\documentclass[twoside,a4paper,11pt]{article}
\usepackage{graphicx}
\usepackage{color}
\usepackage{array}
\usepackage[official]{eurosym}
\usepackage{multirow}
\usepackage{ekaia}

\begin{document}

 \izenburua{Euskahaldun: euskararen aldeko martxa baten sare sozialetako islaren bilketa eta analisia}

 \begin{autoreak}
  \textit{Arkaitz Zubiaga}
  \linebreak
  University of Warwick
  \linebreak
  arkaitz@zubiaga.org
  \linebreak
 \end{autoreak}

 \datak{2015-05-28}{}

 \begin{laburpena}
  Gutxi dira sare sozialetan oinarrituz euskara landu duten ikerketa-lanak, eta are gutxiago Euskal Herrian ospatutako ekitaldiek sare sozialetan utzitako aztarnak aztertu dituztenak. Hutsune hori bete eta arlo honetan ikerketa sustatzeko asmoz, lan aitzindaria aurkeztea du helburu artikulu honek. Horretarako, ``Euskahaldun'' lemapean 2015eko Korrika martxak Twitter sare sozialean sortutako jarduna batzeko jarraitutako metodologia azaldu eta emaitza aztertzen dugu artikulu honetan. Gure analisiak erakusten duenez, emozio handieneko momentuak Twitterren ere islatzen dira, txio kopuru handiagoa sortuz. Horrez gain, euskal komunitatean ikusgarritasuna lortu eta informazioa lau haizetara zabaltzeko ekitaldiarekin lotutako kontu ofiziala izatearen garrantzia erakusten dugu, eta baita kazetari eta komunikabideen parte-hartzearen beharra ere. Guztion eskura jarri ditugu Twitterrekin antzeko analisiak egiteko tresnak, antzeko ikerketa lanak sustatu asmoz.
 \end{laburpena}

 \begin{hitz-gakoak}
  sare sozialak, ekitaldiak, Twitter, jarrera, datu meatzaritza
 \end{hitz-gakoak}

 \begin{abstract}
  This work is motivated by the dearth of research that deals with social media content created from the Basque Country or written in Basque language. While social fingerprints during events have been analysed in numerous other locations and languages, this article aims to fill this gap so as to initiate a much-needed research area within the Basque scientific community. To this end, we describe the methodology we followed to collect tweets posted during the quintessential exhibition race in support of the Basque language, Korrika. We also present the results of the analysis of these tweets. Our analysis shows that the most eventful moments lead to spikes in tweeting activity, producing more tweets. Furthermore, we emphasise the importance of having an official account for the event in question, which helps improve the visibility of the event in the social network as well as the dissemination of information to the Basque community. Along with the official account, journalists and news organisations play a crucial role in the diffusion of information. In order to encourage others to perform further research in the field, we make all the tools publicly available.
 \end{abstract}


\section{Sarrera}
 
Interneten garapenarekin batera tresna berriak garatu eta argitaratu dira, Web 2.0 fenomenoaren garrantzia areagotuz. Izan ere, Web 2.0 delakoan erabiltzaileek, Internetetik informazioa jasotzeaz gain, modu errazean egin dezakete ekarpena eta sareko informazio-jarioa eta ezagutza aberastu. Web 2.0 tresnen artean, sare sozialak dira nabarmendu daitezkeen tresna horietako bat. Sare sozialetan erabiltzaileek parte hartu eta ekarpenak egiteaz gain, lagun-sare bat sortzeko aukera dute, eta ondorioz, aukeratutako lagun-sarearekin informazioa, argazkiak eta bideoak konpartitu, eztabaidatu eta solasteko aukera eskaintzen diete.

Gaur egungo sare sozial nagusien artean, Twitter\footnote{http://twitter.com/} dugu ikerketarako erabiliena. Izan ere, sare sozial honetan erabiltzaileek idazten dituzten mezuak eskuratzeko aukera eskaintzen zaigu, eta horrek aukera paregabea eskaintzen du tresna informatikoen bidez datu-bildumak batu eta ikertu ahal izateko. Beste sare sozial batzuetan ez bezala, Twitterren gehienez 140 karaktereko mezuak bidaltzera mugatuta daude erabiltzaileak. Mezu labur hauek txio izena dute, eta interesgarri iruditzen zaion orok zabaldu egin dezake; beste norbaiten txioa zabaltze horri bertxio deritzo. Erabilgarritasunaren aldetik abantaila paregabeak eskaintzen ditu horrek. Besteak beste, oso erraza da edonondik sakeleko telefonoa erabiliz zure begien aurrean ikusten ari zarenaren gainean txiokatzea. Horrek, esate baterako, herri kazetaritzaren hazkundea ahalbidetu du, edonork informa baitezake edonondik, eta Twitter bezalako sare handi batean ikusgarritasuna lortu. Era berean, ekitaldi bat ospatzen ari den bitartean, erabiltzaileek iritziak, argazkiak, bideoak, eta albisteak zabal ditzakete.

Twitterrek datuok biltzeko eskaintzen dizkigun aukerei esker, interes handia piztu da komunitate zientifikoan, ikerketa egiteko informazio iturri aberatsa baita. Ondorioz, asko dira Twitterren ekitaldi ezberdinek utzitako islaren ikerketa egin duten lanak, tartean New Yorkeko Occupy Wall Street \cite{conover2013digital,gleason2013occupy}, Arabiar Udaberria \cite{lotan2011arab} edo Munduko Futbol Txapelketa \cite{godfrey2014case,yu2015world} bezalakoak analizatuz. Baina ekitaldi horiek guztiak erdaraz eta nazioartean gertatutakoak dira, eta orain arte ez da antzeko lanik egin euskaraz zein Euskal Herrian gertatutako ekitaldiekin.

Euskal komunitate zientifikoa arlo honetan lan gehiago burutu eta Euskal Herriko ekitaldiak aztertzera animatu nahian, lan aitzindaria aurkezten dugu artikulu honetan. Hemen aurkezten den lana IkerGazte kongresuan\footnote{http://www.ueu.eus/ikergazte/} emandako \textit{``Sare sozialetatik erauzitako datuetan oinarritutako ikerketa''} tailerraren prestaketaren emaitza da. Twitter iturri bezala hartuta, ikertzea helburu izanez datuak biltzeko metodologia azaltzen dugu, horretarako behar diren tresnak guztion eskura jarriz. Euskararen kasua aztertzeko aukera ezinhobea eskaintzen duen ekitaldi baten adibidea aurkezten dugu artikuluan, Korrikarena\footnote{http://www.korrika.eus/} hain zuzen ere, analisia egin eta emaitzak azalduz.

Metodologia hau definituz eta kodea guztion eskura jarriz euskal komunitate zientifikoak arlo honetan ikertzen jarraitzea dugu helburu nagusi lan honekin, euskarazko eta Euskal Herriko ekitaldiek sare sozialetan sortzen duten isla aztertuz euskal komunitatearen jarrera hobeto ulertzeko.

\section{Twitterren gaineko ikerketa}
 
Twitter sare sozial nagusienetakoa bilakatu da gaur egun. Twitterrek berak ematen dituen estatistiken arabera\footnote{https://about.twitter.com/company}, 300 milioi erabiltzailetik gora dira hilero ekarpenen bat egiten dutenak, eta egunero 500 milioi txioren inguruan jasotzen ditu sare sozialak. Zaila da zehatz-mehatz txio hauen guztien artean euskarazkoak zenbat diren jakitea, baina Umap-ek\footnote{http://umap.eu/}, euskarazko jarduna hein handi batean batzen duen tresnak, ematen ditu gutxi gorabeherako datu batzuk. Horien arabera\footnote{http://umap.eu/media/pdf/umap\_2015\_4.pdf}, 7.000 erabiltzaile baino gehiago dira nagusiki euskaraz txiokatzen aritzen direnak, eta 2015eko apirilean 226.000 txio inguru ziren euskarazkoak, egunean 7.539 batez beste. Horrenbesteko datu-jario etengabea duen informazio-iturriak ikerketarako bilduma paregabea eskaintzen du ondorioz, eta asko dira horri etekina ateratzen ari direnak.

Azken urteotan, Twitter sare sozialean elkarbanatutako txioen gainean egindako ikerketa lanak nabarmen hazten ari dira. Sare soziala 2006ko martxoan\footnote{https://about.twitter.com/milestones} abiatu bazen ere, 2010. urtean argitaratu ziren lehen ikerketa-lanak \cite{kwak2010twitter}. Orduz geroztik, etengabe hazten doa Twitter sare sozialean oinarritzen diren ikerketa-lanen kopurua. Izan ere, Twitter sare sozialak datuak eskuratu ahal izateko aukera paregabeak eskaintzen dizkigu ikertzaileoi, doan eta nahiko erraz lor baitaitezke datu bilduma handiak. Twitterrek hiru abantaila nagusi eskaintzen dizkigu ikerketa dugunean helburu:
 
\begin{itemize}
 \item \textbf{Datu-bilduma handiak lortzeko erraztasuna:} Twitter APIa baliatuz datuak eskuratzea doakoa da, eta muga batzuk dituen arren (esate baterako, eskaera kopurua mugatua du, eta ez du datu-base osoaren \%1 baino gehiago eskuratzen uzten), datu-bilduma handiak lortzeko aukera eskaintzen du, eta normalean muga horiek ez dira oztopo izaten, datu-bilduma erraldoiak nahi baldin ez badira behintzat.

 \item \textbf{Datuen naturaltasuna:} beste ikerketa-metodo tradizional batzuekin alderatuta, Twitterretik jasotako datuak naturalagoak dira. Metodo tradizionalean parte-hartzaileak batu ohi dira, eta ikerketaren helburua azaldu ondoren, haien erantzunak zein bestelako datuak gordetzen dira. Horrelako kasuetan gerta litekeena da parte-hartzaileek emandako erantzunak guztiz egiazkoak ez izatea, eta, ondorioz, ikerketaren emaitza nolabait izorratzea. Twitter bezalako sistemen abantaila da erabiltzaileek modu naturalean idazten dutela, nahi dutelako, eta inork behartu gabe.
 
 \item \textbf{Mota ezberdinetako datuak lor daitezke:} Askotan Twitter lagunarteko elkarrizketetarako eta ``orain zer egiten ari naizen'' bezalakoak zabaltzeko baino erabiltzen ez dela uste den arren, beste mota askotako informazioa topa daiteke bertan. Erabiltzaile askok eta askok inguruko ekitaldiez hitz egin ohi dute, eta batzuetan, azken orduko albisteak gertatu ahala, bertan dauden lekukoek informazio esklusiboa ere eman izan dute, tresnaren ezaugarriei esker.
\end{itemize}
 
Goiko abantailei erreparatuz ikerketarako informazio-iturri perfektua dirudien arren, baditu kontuan izan beharreko hainbat desabantaila ere:

\begin{itemize}
 \item \textbf{Datuak eskuratzen azkar ibili beharra:} Twitter APIa doakoa den arren, muga nagusienetakoa da datu historikoak eskuratzeko zailtasuna. Twitterren APIa bereziki prestatuta dago denbora errealean konpartitzen diren txioak, zein azken astean bidali diren txioak eskuratzeko, eta txio zaharragoak eskuratu nahi ditugunean arazoak izan ditzakegu askotan. Txioek Twitterren datu-basean jarraitzen duten arren, ez daude modu errazean eskuragarri. Horrelakoetan, ordainpeko zerbitzuetara jotzea gomendatzen da; baina hori oztopo handia izan daiteke askotan ikertzaileontzako. Ondorioz, adi egotea eta txioak denbora errealean edo berehala, ordu gutxi batzuen buruan, eskuratzea da egokiena.
 
 \item \textbf{Hizkuntza informala:} Twitterren konpartitzen den edukia aztertzean ez dute ohiko analisi linguistikorako tresnek ondo funtzionatzen. Arazo nagusia da sare sozialetan hizkuntza informala erabili ohi dela, eta, beraz, akats ortografikoez eta laburdurez josita egon ohi dela. Hori dela eta, zailagoa da edukia aztertzea, azken urteotan alor horretako ikerketa hobetzen ari den arren.
 
 \item \textbf{Gezurrak:} Albisteak jarraitzean, esate baterako, kontu handiz hartu behar izaten da Twitterren irakurritakoa, oso erraza baita albisteak asmatu eta gezurrezko istorioak sarean jartzea. Horrek ikerketan ere eragina izan dezake \cite{zubiaga2014tweet}, eta, beraz, kontuz ibili behar da sare sozialetako edukiak aztertzean.
 
 \item \textbf{Datu demografikoen gabezia:} Twitterreko erabiltzaileak aztertzean, litekeena da beraien datu demografikoak erabili nahi izatea batzuetan, lagina hobeto aukeratzeko edo laginaren ezaugarri demografikoak jakiteko. Zoritxarrez, baina, Twitterreko erabiltzaileek ez dute datu demografikorik argitaratzen beren profilean, eta ezinezkoa da erabiltzaile baten adina, sexua, eta abar jakitea. Kokapena jakin ahal izaten da batzuetan, baina kokapenari dagokion eremuan erabiltzaileak edozer jar dezakeenez, ez da oso fidagarria.
 
 \item \textbf{Demografikoki adierazgarri ez izatea:} Twitterren dauden erabiltzaileek ez dute gizartearen banaketa demografikoa guztiz islatzen. Sloan et al.-ek \cite{sloan2015tweets} erakutsi zutenez, Twitterren dauden erabiltzaileak gazteak dira batez ere, eta haien profilak ez du gizartearekin guztiz bat egin.
\end{itemize}

Desabantaila horiek guztiak kontuan izan beharrekoak dira, batez ere Twitterretik datuak biltzeko lagina ondo aukeratuz, datuen bilketan alborapenik egon ez dadin. Horrez gain, euskararekin lan egitean aipatu beharreko beste desabantaila bat da Twitterrek eskaintzen duen hizkuntza iragazkia; izan ere, posible da txioak batzerakoan zein hizkuntzakoak jaso nahi diren zehaztea. Zoritxarrez, ordea, Twitterrek gaur egun ez du euskara antzemateko gaitasunik, eta, beraz, ezin zaio esan euskarazko txioak bakarrik jaso nahi ditugula. Oztopo horiek alde batera utzi eta ahal den neurrian saihestuta, eta abantailei erreparatuz, ordea, euskaraz sare sozialen gainean dagoen ikerketa areagotu egin behar dela uste dugu.

Lan honetan aurkezten den analisiari dagokionez, gainera, goiko desabantaila horiek ez dute eragin handirik, eta etorkizunean landu beharreko erronka gisa uzten ditugu. Era berean, hemen aurkezten dugun hurbilketa beste hizkuntza batzuetan eta orokorrean komunitate zientifikoan onartuta dagoena da. Artikulu honetan, Korrikaren gainean zabaldutako txioen gaineko analisia eginez, ez dugu gizartea islatzen duenik aldarrikatu nahi; beste ikuspuntu batetik, euskarazko ekitaldi batek sare sozialetan eta bereziki Twitterren sortzen duen isla aztertu nahi dugu, gaur egun sareko presentziak adierazten baitu, hein handi batean, komunitate eta ekitaldi zehatz baten ikusgarritasuna.

Ekitaldiak Twitter iturri gisa hartuta ikertzeari dagokionez, lan gehienak ingelesez idatzi dira, eta ingelesezko txioak aztertu dituzte \cite{conover2013digital,gleason2013occupy,lotan2011arab,godfrey2014case,yu2015world}. Beste hizkuntza batzuetan, gaztelaniaz \cite{rodriguez2011redes} zein portugesez \cite{da2014discurso} esate baterako, argitaratu da ikerketa-lanen bat. Euskaraz, ordea, hutsunea sumatu dugu arlo honetan, eta horixe betetzea da lan honen helburu nagusia. Bizkarguenagak \cite{bizkarguenaga2012euskal}, esate baterako, ikuspegi soziologikotik aztertu zuen gazteek sare sozialetan nola sortzen duten euskal identitatea, Facebook sare soziala aztertuz. Lan honek, ordea, gizarte-zientzien ikuspegi horretatik analisi kualitatiboa eskaintzen du, beste ikuspegi teknikoago batetik eskain daitekeen analisi kuantitatiboari gehiegi erreparatu gabe.

Euskaraz dugun hutsunea bete nahian, lan honek Twitter sare sozialean ekitaldi baten gainean zabaldutako mezuak modu kuantitatiboan aztertzeko behar den metodologia definitzen du, beharrezko tresnak ikertzaileen eskura jarriz, eta 2015ean ospatutako Korrika ekitaldiaren gaineko analisia aurkeztuz.

\section{Datu bilketa}
 
Bi erronka nagusi ditu Twitter sare sozialean zabaltzen diren mezuak jarraitu eta ikerketa egin ahal izateko artxibatzeak: datuak batuko dituen tresna prestatzea, eta, Twitterren jarraitu nahi den hitz eta traol sorta prestatu ondoren, datu-bilketa egitea.

\subsection{Datuak biltzeko tresna}

Twitterren APIak bi zerbitzu ezberdin eskaintzen ditu: (1) REST APIa, azken egunetako txioak eskuratzeko, eta (2) streaming APIa, txioak denbora errealean eskuratzeko. Kasu honetan bigarrena erabili dugu, streaming APIa; izan ere, hainbat egun irauten duen ekitaldi baten jarraipena egiteko, errazagoa izaten da aukera hau erabiltzea.

Python programazio-lengoaia baliatu dugu datu-bilketarako. Lengoaia honek badu txioak batzeko pakete bat, lana asko errazten duena, tweepy\footnote{http://www.tweepy.org/} izenekoa. Twitterretik txioak eskuratu eta gordetzeko garatutako kodea github-en aurki daiteke\footnote{https://github.com/azubiaga/twitter-tools/tree/master/tweet-collection}. Hori erabiltzeko, bete beharreko lehen pausoa Twitterren aplikazioa sortzea da, \textit{http://apps.twitter.com/} helbidean. Ondoren, Twitterrek emango dizkigun lau kodeak \textit{twitter.ini} fitxategian kopiatu behar dira. Hori egin ondoren, \textit{tweetcollector.py} programaren bitartez txioak batzen hasi ahal izango dugu, honako komandoa erabiliz:

\vspace{0.5cm}

\texttt{python tweetcollector.py bilketa-mota ekitaldi-izena bilaketa-katea}

\vspace{0.5cm}

\noindent Non:

\begin{itemize}
 \item \textit{bilketa-mota} hauetako bat izan daitekeen: 'search-recent' (azken txioak eskuratzeko), 'search-popular' (azken asteko txio garrantzitsuenak eskuratzeko) edo 'stream' (txioak denbora errealean eskuratzeko).
 \item \textit{ekitaldi-izena} eskuratu nahi den ekitaldiaren izena izango den. Nahi den izena aukera daiteke hemen, eta programak izen hori duen karpeta sortuko du \textit{data} karpetaren barruan, non txioak gordeko diren.
 \item \textit{bilaketa-katea} Twitterren egin nahi dugun bilaketa den, hitz gako ezberdinak hutsunez banandurik.
\end{itemize}
 
\subsection{Datu-bilduma: Korrika 2015}
 
\textit{tweetcollector.py} programa baliatuz, Korrika ospatu zen 11 egunetan zehar egin genuen datu-bilketa, hau da, 2015eko martxoaren 19tik 29ra. Bilaketa-katea Korrikarekin lotutako hainbat traolekin osatu genuen, tartean \#korrika, \#euskahaldun, \#korrika2015, \#korrika15 eta \#korrikazuzenean. Horrenbestez, honako komandoarekin abiarazi genuen programa, eta 11 egun horietan zehar martxan mantendu:

\vspace{0.5cm}

\texttt{python tweetcollector.py stream korrika15 \char`\"\#korrika\char`\" \char`\"\#euskahaldun\char`\" \char`\"\#korrika2015\char`\"...}

\vspace{0.5cm}

Datu-bilketaren emaitza 38.276 txioko bilduma izan zen. Hurrengo atalean, txio-bilduma horren gaineko analisia aurkezten dugu. Datu-bilduma hau sarean argitaratu dugu\footnote{https://github.com/azubiaga/korrika15}. Datu-bilduma honekin batera jaits daitekeen \textit{korrika.json} fitxategia erabiltzen dugu analisirako iturri moduan, zeinak 38.276 lerro dituen, lerro bakoitzean txio bana, JSON formatuan. JSON formatuak dituen eremuen gaineko xehetasun gehiago Twitterren webgunean bertan aurki daiteke\footnote{https://dev.twitter.com/overview/api/tweets}.
 
\section{Analisia}

Atal honetan analisia aurkezten dugu, bi zatitan banatuta: edukiaren analisia eta erabiltzaileen analisia. Analisirako landu ditugun programak eskuragarri daude sarean github-en\footnote{https://github.com/azubiaga/twitter-tools}. Horiez gain, Barber\'ak \cite{barbera2015birds} garatutako beste programa batzuk ere erabili ditugu\footnote{https://github.com/pablobarbera/pytwools}.
 
\subsection{Edukiaren analisia}

Hasteko, edukiari erreparatuz egingo dugu analisia. Korrikak hamaika egunez jardun zuen Euskal Herrian zehar lasterka, eta aztertu nahiko genuke sare sozialetan utzitako isla nola sakabanatu den denboran zehar. Zein momentutan txiokatu zuen gehien jendeak? Horretarako, txio kopuruaren histograma aztertuko dugu, \textit{histogram.py} programa baliatuz, eta honako komandoaren bitartez:

\vspace{0.5cm}

\texttt{python histogram.py korrika.json korrika-histograma.dat h}

\vspace{0.5cm}

\begin{figure}[tbh]
 \begin{center}
  \includegraphics[width=0.60\textwidth]{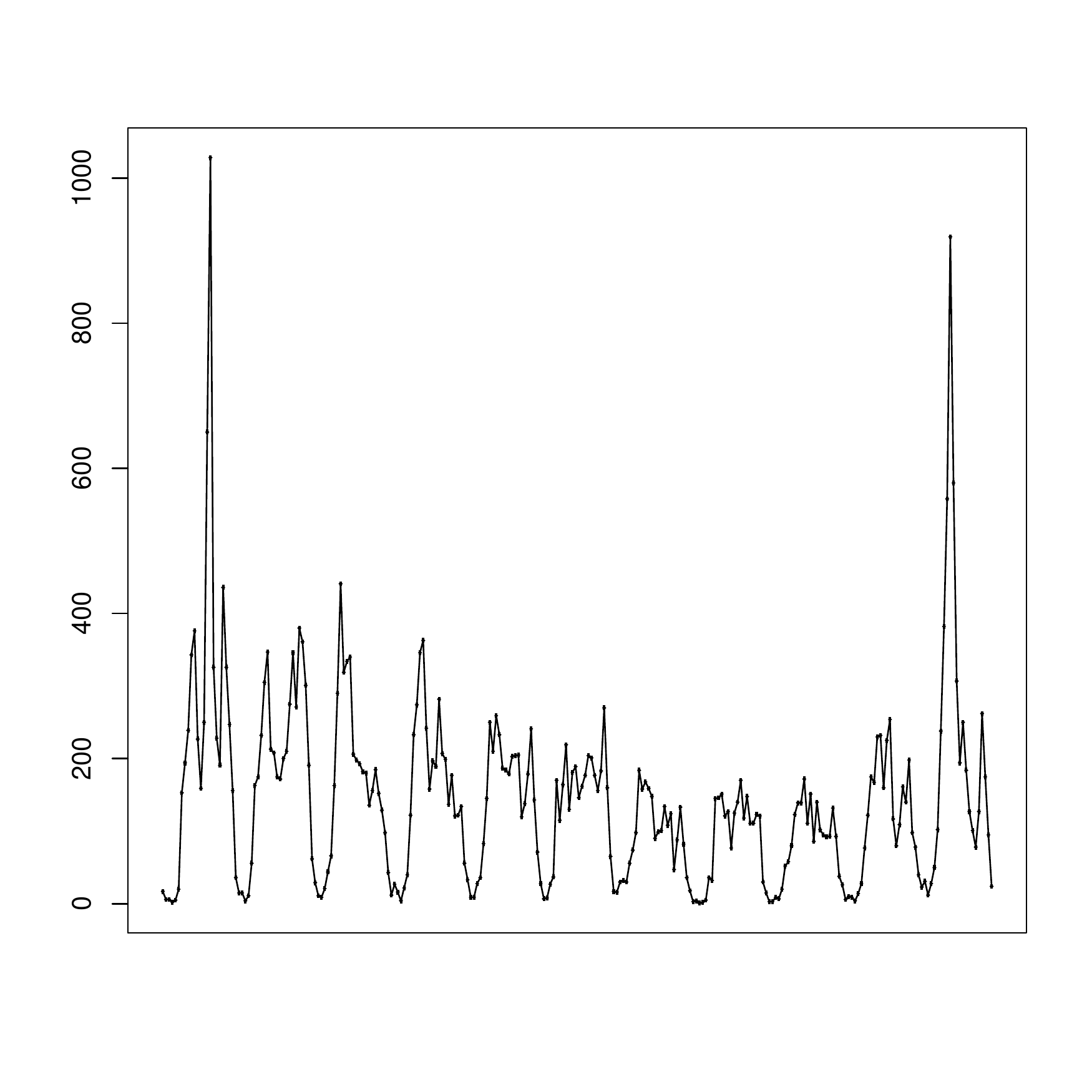}
  \caption{Txioen histograma, orduro bidalitako txio kopurua erakusten duena.}
  \label{fig:histograma}
 \end{center}
\end{figure}

Emaitza \ref{fig:histograma}. irudian ikus daiteke. Hainbat gauza erakusten ditu histograma honek. Argi dagoena da, gauetan, Korrikak aurrera diharduen arren, txio kopuruak izugarri egiten duela behera; hori dela eta, hori dela eta histogramak 11 egunei dagozkien igoera eta jaitsiera nabarmenak ikus daitezke. Horrez gain, lehen eta azken eguna izan ziren txio gehien bidalitakoak; momenturik txiokatuena hasierakoa izan zen, ordubetean 1000 txio baino gehiagorekin. Ondorengo egunetan, txio kopurua gutxika-gutxika jaitsi zela dirudi, Korrikaren kontaketa ez baita hain berria eta interesa nolabait jaitsi zelako seguruenik; eta azkenik, nabarmen egiten zuen gora berriz azkenengo egunean, Bilboko helmugara iristearekin batera.

Edukiari gertuagotik erreparatuz, interes handiena piztu zuten txioak zein izan ziren jakin nahi dugu. Horretarako, bertxio kopurua aztertzen dugu. Zein izan ziren bertxio gehien jaso zituzten txioak? Horretarako, \textit{top-tweets.py} programa erabiliz, txio bertxiotuenen rankinga sortu dugu:

\vspace{0.5cm}

\texttt{python top-tweets.py -v retweets -f korrika.json -k 10}

\vspace{0.5cm}

\begin{table}[hbt]
 \footnotesize
 \begin{center}
  \begin{tabular}{| l | l |}
   \hline
   \textbf{Erabiltzailea} & \multirow{2}{*}{\textbf{Txioa}} \\
   \textbf{(Bertxioak)} &  \\
   \hline
   \hline
   @MeriLing1 & Behin batean herri txiki handi bateko milaka pertsonek, zahar eta \\
   (214) & gaztek gau ta egun korrika egin zuten euren hizkuntzaren alde \#mikroipuina \\
   \hline
   @euskaltegia & Herri bat, euskaraz bizi nahian, tinko, harro. 92 urteko emakumea, \\
   (170) & Arraiotz-en. ZORAGARRIA \#KorrikaZuzenean http://t.co/VsIcWoQhhM \\
   \hline
   @fm914 & Ederra Indar Gorrikoek \#Korrika-ri egindako agurra. \\
   (149) & Argazki andana gure guaxapean. Esker mila! http://t.co/pO3IpvLCbP \\
   \hline
   @ZuriHesian & Hunkigarria da pentsatzea edozein ordutan, nonbaiten, zoro zoragarri \\
   (145) & batzuk euskararen alde \#korrika dabiltzala :) http://t.co/X9OEzXoq8p \\
   \hline
   @maia\_jon & Korrika edo Giza katea bezalako ekimenek sortzen duten energia \\
   (132) & kolektiboa ez al dira nahikoa froga herri honen bidea zein den ohartzeko? \\
   \hline
   @korrika\_aek (126) & Hasi da korrika19 !! Tipi tapa tipi tapa KORRIKA!! \\
   \hline
   @Hostinet & \'Unete a la \#Korrika y contrata tu \#dominio .eus y \#hosting web por \\
   (98) & tan solo 29,95 \euro{} con @Hostinet y @puntueus https://t.co/RTxoGkyKOn \\
   \hline
   @garesko\_auzalan (96) & Korrikaren gazte kilometroa Garesen! http://t.co/HHPjaVPqJ8 \\
   \hline
   @Lupilakasta (91) & Fiterotik Bilbora abia dadin Korrika!! http://t.co/w4esPCX3Xw \\
   \hline
   @EuskalakariAEK & 2.500 km egin ostean, lekukoa Bilbora iritsi da. Hona hemen mezua! \\
   (89) & Gora \#korrika! Eskerrik asko, @katuajea! http://t.co/yB5C1FcWk4 \\
   \hline
  \end{tabular}
  \caption{Txio bertxiotuenak, bertxio kopuruaren arabera ordenatuta.}
  \label{tab:txio-bertxiotuenak}
 \end{center}
\end{table}

Eta \ref{tab:txio-bertxiotuenak}. taulan zerrendatu ditugu Korrikan zehar bertxio gehien jaso zituzten 10 txioak. Mota ezberdinetako txioak ikus ditzakegu zerrenda horretan, baina, batez ere, ekitaldiarekin lotutako gertakizunak kontatzen dituzten txioak (adibidez, ``Hasi da korrika19!'') eta txio aldarrikatzaileak (adibidez, ``Korrika edo Giza katea bezalako ekimenek sortzen duten energia kolektiboa ez al dira nahikoa froga herri honen bidea zein den ohartzeko?'') aurki ditzakegu. Euskarazko 9 txio izateaz gain, gaztelaniazko bat ere lehen 10 hauen artean dago; izan ere, .eus Interneteko domeinuak deskontupean eskuratzeko promozio batek bertxio asko jaso zituen. Aurrez komentatu dugun bezala, momentuz, soluzio errazik gabeko erronka da Twitteren hizkuntza automatikoki antzeman ahal izatea.

Edukia beste ikuspuntu batetik begiratuta, txioak nondik bidali diren jakin nahi dugu. Zoritxarrez, txio guztiek ez dute eskuragarri geokokapena, eta ezin da beti jakin txio bat nondik bidali den. Hori dela eta, geokokapena eskaintzen duten txioetara mugatu behar dugu ondoko analisia. Txioetatik geokokapena dutenak erauzteko, \textit{coordinates.py} programa erabiltzen dugu, behean azaldutako komandoaren bitartez, eta ondoren CartoDB\footnote{http://www.cartodb.com/} doako zerbitzua erabiliz bistaratu. CartoDB zerbitzuan erregistratu beharra dago, baina guztiz doakoa da, eta nahikoa da \textit{coordinates.py} programak sortzen duen koordenatu-zerrenda igotzearekin.

\vspace{0.5cm}

\texttt{python coordinates.py korrika.json korrika-koordenatuak.dat}

\vspace{0.5cm}

\begin{figure}[tbh]
 \begin{center}
  \includegraphics[width=0.60\textwidth]{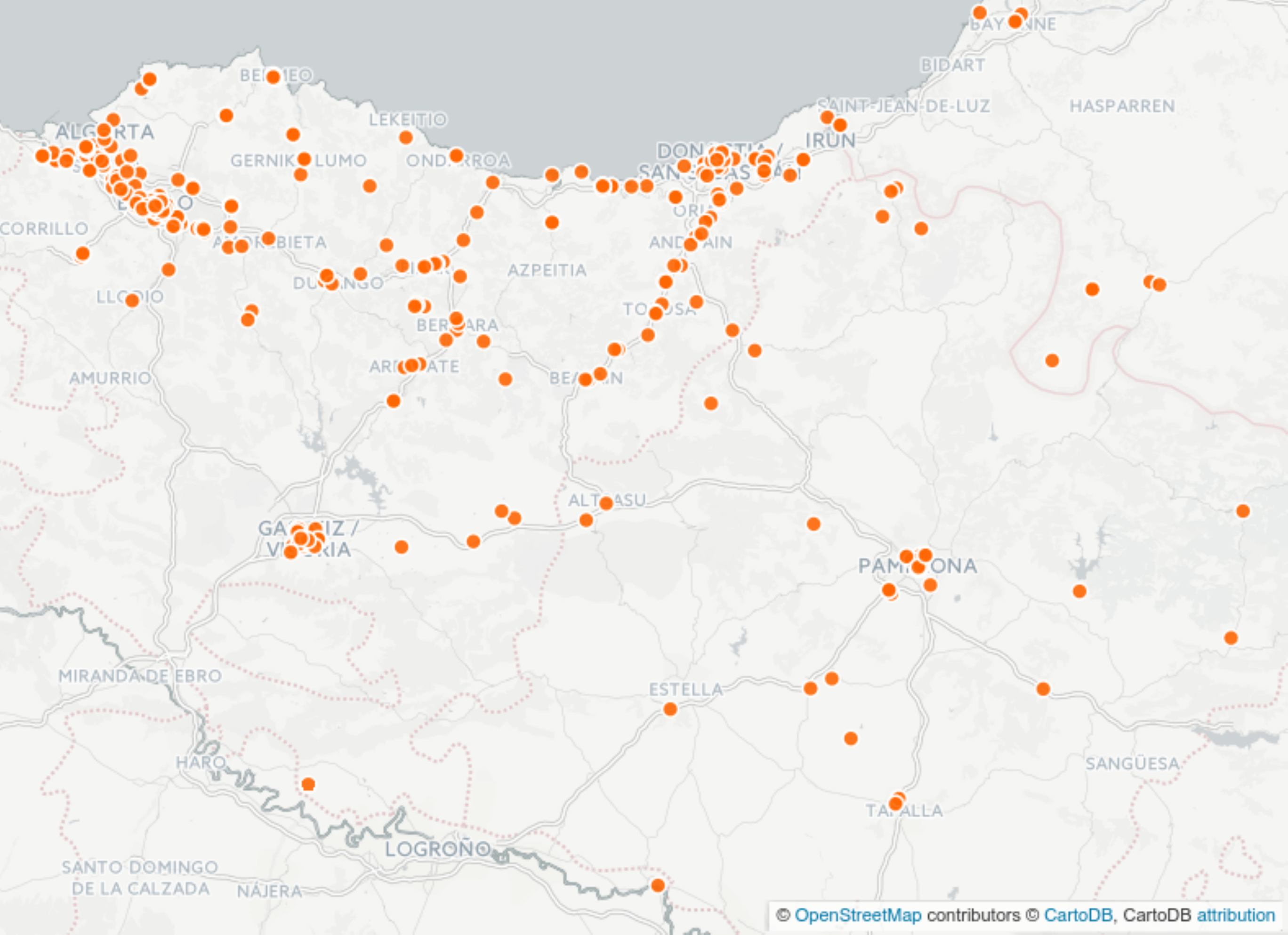}
  \caption{Txio geokokatuen mapa.}
  \label{fig:geokokatuak}
 \end{center}
\end{figure}

Txio geokokatuekin sortutako mapa \ref{fig:geokokatuak}. irudiak erakusten du. Mapa horrek erakusten duenez, txioak Euskal Herri osoan zehar zabaldu ziren, lurralde guztietan. Hiriburuetan eta inguruetan txio gehiago dauden arren, Arrasate, Durango, Gernika, Tolosa, Altsasu, Lizarra, Tafalla eta beste herri askotatik ere hainbat txio bidali ziren.
 
\subsection{Erabiltzaileen analisia}

Edukia aztertu ondoren, sare sozialetan garrantzi handikoa den beste faktore bat aztertzeari ekingo diogu; erabiltzaileak, alegia. Aktiboenen zerrenda sortuz hasiko dugu erabiltzaileen analisia: nork bidali zituen txio gehien Korrikaren gainean? Horretarako, \textit{top-tweets.py} programa erabil dezakegu, komando honen bitartez:

\vspace{0.5cm}

\texttt{python top-tweets.py -v users -f korrika.json -k 10}

\vspace{0.5cm}

\begin{table}[hbt]
 \small
 \begin{center}
  \begin{tabular}{| l | r |}
   \hline
   \textbf{Erabiltzailea} & \textbf{Txio kopurua} \\
   \hline
   \hline
   @idorrokia & 1085 \\
   \hline
   @EuskalakariAEK & 910 \\
   \hline
   @korrika\_aek & 614 \\
   \hline
   @HamaikaTb & 444 \\
   \hline
   @EAPortugalete & 239 \\
   \hline
   @naiz\_info & 233 \\
   \hline
   @gaztea & 227 \\
   \hline
   @berria & 216 \\
   \hline
   @euskalirratiak & 208 \\
   \hline
   @anabarri72 & 204 \\
   \hline
  \end{tabular}
  \caption{Erabiltzaile aktiboenak.}
  \label{tab:erabiltzaile-aktiboenak}
 \end{center}
\end{table}

Korrikan zehar txio gehien bidali zituzten 10 erabiltzaileak \ref{tab:erabiltzaile-aktiboenak}. taulak erakusten dira. Zerrenda honetan, Korrikarekin lotutako kontuak (@EuskalakariAEK eta @korrika\_aek) eta komunikabideak (@HamaikaTb, @naiz\_info, @gaztea, @berria, @euskalirratiak) dira nagusi, baina badaude erabiltzaile arrunten (@idorrokia, @anabarri72) eta alderdi politikoen (@EAPortugalete) kontuak ere tartean. Dena den, zerrenda honek txio kopurua baino ez du erakusten, eta sakonago aztertu nahiko genuke nor izan diren interesgarrienak komunitatearentzat. Hori neurtzeko, bertxio kopuruari erreparatuko diogu. Zein erabiltzailek jaso zuten bertxio gehien? \textit{top-users.py} programa erabiliz, zerrenda hau eskura dezakegu:

\vspace{0.5cm}

\texttt{python top-users.py korrika.json}

\vspace{0.5cm}

\begin{table}[hbt]
 \small
 \begin{center}
  \begin{tabular}{| l | r |}
   \hline
   \textbf{Erabiltzailea} & \textbf{Bertxio kopurua} \\
   \hline
   \hline
   @EuskalakariAEK & 1642 \\
   \hline
   @naiz\_info & 918 \\
   \hline
   @HamaikaTb & 720 \\
   \hline
   @hirual & 640 \\
   \hline
   @korrika\_aek & 568 \\
   \hline
   @argia & 482 \\
   \hline
   @berria & 385 \\
   \hline
   @larbelaitz & 363 \\
   \hline
   @gaztea & 305 \\
   \hline
   @MeriLing1 & 243 \\
   \hline
  \end{tabular}
  \caption{Erabiltzaile bertxiotuenak.}
  \label{tab:erabiltzaile-bertxiotuenak}
 \end{center}
\end{table}

Hain zuzen, \ref{tab:erabiltzaile-bertxiotuenak}. taulak erakusten du bertxio gehien jaso zituzten 10 erabiltzaileen zerrenda. Lehenago azaldutako erabiltzaile aktiboenekin alderatuta, badaude hor goian mantendu diren batzuk, baina beste batzuk aldatu egin dira. Korrikarekin lotutako kontuak hor goian daude oraindik (@EuskalakariAEK, @korrika\_aek), eta baita aurretik genituen hainbat komunikabide ere (@naiz\_info, @HamaikaTb, @berria, @gaztea); baina oraingoan beste komunikabide bat ere agertu zaigu tartean (@argia), hainbeste txiokatu gabe bertxio kopuru handia lortu duena. Oraingoan, baina, erabiltzaile arruntak desagertu egin dira zerrendatik, eta hainbat kazetarik hartu dute haien lekua (@hirual, @larbelaitz, @MeriLing1). Horren arabera, badirudi kazetariek gutxiagotan txiokatu dutela, baina interes handiagoa piztuz.

Erabiltzaileen analisi honetan sakondu eta grafikoki ikusi ahal izateko, erabiltzaile nagusien sare soziala bistaratu nahi dugu, amaitzeko. Horretarako, erabiltzaileen arteko interakzioak erauziko ditugu lehenengo. Hau da, noren txioak bertxiotu edo erantzun ditu erabiltzaile bakoitzak? Horretarako, @a erabiltzaileak @b erabiltzailearen txio bat bertxiotzen badu, interakzio moduan hartuko dugu kontuan. Era berean, @a erabiltzaileak @b-ren txio bati erantzuten badio. Korrikako txio guztietatik interakzio horien zerrenda erauzteko, \textit{interactions.py} erabiliko dugu, honako komandoarekin:

\vspace{0.5cm}

\texttt{python interactions.py korrika.json korrika-interakzioak.csv}

\vspace{0.5cm}

\begin{figure}[ptbh]
 \begin{center}
  \includegraphics[width=0.90\textwidth]{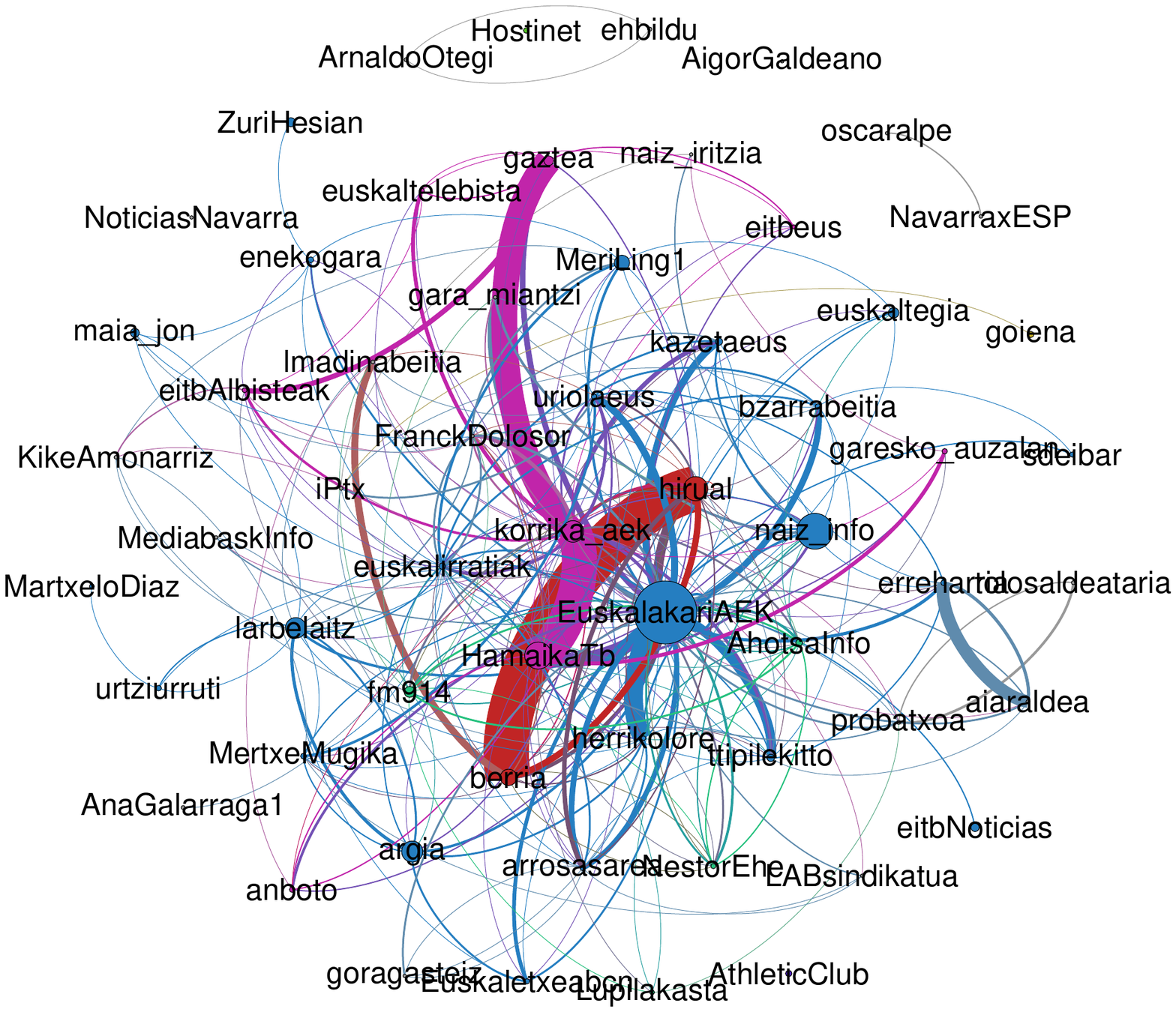}
  \caption{Erabiltzaile esanguratsuenekin osatutako grafoa.}
  \label{fig:grafoa}
 \end{center}
\end{figure}

Ondoren, interakzioen fitxategi hau bistaratzeko Gephi\footnote{http://gephi.github.io/} beharko dugu. Gephi software librea da eta plataforma ezberdinetarako (Windows, Linux, Mac) dago eskuragarri. Sortu berri dugun \textit{korrika-interakzioak.csv} fitxategia Gephi-n kargatu dezakegu orain, eta sare soziala bistaratu.

Emaitza \ref{fig:grafoa}. irudian ikus daiteke. Erabiltzaile nabarmenenak bistaratzen dira bertan, eta beraien arteko loturaren sendotasunak interakzio kopurua adierazten du; koloreek, berriz, komunitateak. Grafo honek @EuskalakariAEK kontu ofizialaren garrantzia nabarmentzen du, Korrikaren sareko presentzia eta informazioaren zabalpenerako ezinbestekoa izan zena. Etorkizunean ospatuko diren ekitaldientzako lezioa ere ematen digu honek, kontu ofizial bat edukitzea zenbateraino den garrantzitsua erakusten baitu. Kontu ofizialaz gain, irudiak erakusten digu inguruko erabiltzaileak gehienbat kazetariak eta komunikabideak direla. Datu horrek kazetari eta komunikabideen garrantzia erakusten du, ekitaldiaren ikusgarritasuna areagotu eta informazioa zabaltzeko lanetan. Horiez gain, beste erabiltzaile mota batzuen garrantzia ere ikus dezakegu, tartean futbol taldeak (@sdeibar, @AthleticClub), musikariak (@ZuriHesian) eta alderdi politikoak (@ehbildu).
 
\section{Ondorioak}

Artikulu honetan Twitter sare sozialean mundu errealeko ekitaldi batek utzitako isla ikertzeko metodologia definitu, azaldu eta praktikan jarri dugu, 2015eko Korrika aztertuz. Orain arte ez dugu antzeko analisirik ikusi euskaraz; hori dela eta, ezinbestekoa iruditzen zaigu, sare sozialen analisiak beste hizkuntza batzuetan bereganatu duen garrantzia euskarara hurbiltzeko. Analisia egiteko tresna guztiak eskuragarri daude sarean, nahi duenak ekitaldi gehiago ikertu ahal izan ditzan eta ikerketa-arlo honetan sakontzeko.

Korrikaren analisia bi zatitan banatu dugu: edukiaren analisia eta erabiltzaileen analisia. Biek ere aurkikuntza interesgarriak egiteko aukera eskaini digute. Edukiari dagokionez, txio kopuruak jasaten dituen gora-beherak azaldu ditugu, non une garrantzitsuenak txio kopuruan islatzen diren, txio gehiago sortuz (kasu honetan, ekitaldiaren hasiera eta amaierako momentuak). Erabiltzaileei dagokienez, ordea, ekitaldiaren zabalkunderako kontu ofiziala izatearen garrantzia nabarmendu dugu. Kontu ofizialaz gain, kazetariak eta komunikabideak ezinbestekoak dira ikusgarritasuna areagotu eta informazioa komunitatean zabaldu ahal izateko. Horiek guztiak kontuan izatea oso garrantzitsua da ekitaldi bat Twitter sare sozialean ikusgarri egin eta erabiltzaileen parte-hartzea sustatzeko.

Korrikaren kasu konkretua azalduz, euskal komunitatea ulertzeaz gain, sare sozialak ikertzeko metodologia eta tresnak elkarbanatuz euskararen inguruko analisi gehiago sustatzea izan da lan honen motibazio eta helburu nagusia.

\section{Etorkizunerako erronkak}

Hemen aurkeztutako analisia hastapenekoa baino ez da, eta etorkizunean antzeko analisi gehiago egin ahal izatea da orain erronka nagusia. Etorkizunean euskal ikerlariek antzeko ikerketa-lanak gehiago lantzea eta analisi gehigarriak eginez euskarazko eta Euskal Herriko gertakizun eta ekitaldiek sare sozialetan sortzen duten islaren gainean gehiago jakitea da helburua.
 
Metodologia zehaztu eta tresnak liberatzeaz gain, euskarazko ekitaldien sare sozialetako isla aztertu eta ulertzeko lan asko dago aurretik oraindik ere. Alde batetik, euskararen inguruko ikerketa-lan gehiago burutzea, eta bestalde, sare sozialek aurkezten dituzten erronkak, lan honetan laburbildu ditugunak, gainditu eta horrelako ikerketa-lanak are gehiago sendotzea. Twitterrek ez duenez ematen batutako txio guztien artean euskarazko txioak zein diren jakiteko aukera, hizkuntza-identifikazio hori egin dezakeen tresna lantzea da etorkizunerako ezinbesteko beharretako bat. Aurretik, horrelako tresna bat lantzeko asmoz, tartean Euskal Herriko Unibertsitatearekin eta Elhuyarrekin landutako TweetLID atazan \cite{zubiaga2014overview} egin genituen lehen saiakerak, eta anotatutako datu-bilduma eskaini genuen, ikertzaileek euskarazko txioen identifikazioa gehiago landu zezaten.
 
% \bibliographystyle{plain}
% \bibliography{zubiaga-ekaia}

\renewcommand\refname{\indent Bibliografia}

\end{document}